\documentclass[journal]{IEEEtran}
\usepackage{cite}
\usepackage[numbers]{natbib}
\usepackage{amsmath,amssymb,amsfonts}
\usepackage{amsthm,amsfonts,amsmath,amssymb}
\usepackage[ruled,vlined]{algorithm2e}
\usepackage{pifont}
\usepackage{subfigure} 
\usepackage{graphicx}
\usepackage{psfrag}

\usepackage{xcolor}
\usepackage{array}
\usepackage{units}
\usepackage{mdframed}
\usepackage{colortbl}
\usepackage{setspace}
\usepackage{textcomp}
\usepackage{multirow}
\usepackage{enumitem}
\usepackage{auto-pst-pdf}
\usepackage{booktabs} % To use tables generate with Excel
\usepackage{steinmetz}
\makeatletter
\newcommand{\mypm}{\mathbin{\mathpalette\@mypm\relax}}
\newcommand{\@mypm}[2]{\ooalign{%
  \raisebox{.1\height}{$#1+$}\cr
  \smash{\raisebox{-.6\height}{$#1-$}}\cr}}
\makeatother

\usepackage[ruled,vlined]{algorithm2e}

\def\BibTeX{{\rm B\kern-.05em{\sc i\kern-.025em b}\kern-.08em
    T\kern-.1667em\lower.7ex\hbox{E}\kern-.125emX}}
\begin{document}

\title{Safe Imitation Learning-based Optimal Energy Storage Systems Dispatch in Distribution Networks}
\author{Shengren Hou, \IEEEmembership{Student Member, IEEE}, Peter Palensky, \IEEEmembership{Senior Member, IEEE},
and Pedro P. Vergara, \IEEEmembership{Senior Member, IEEE}}

% \thanks{This paragraph of the first footnote will contain the date on
% which you submitted your paper for review. It will also contain support
% information, including sponsor and financial support acknowledgment. For
% example, ``This work was supported in part by the U.S. Department of
% Commerce under Grant BS123456.'' }

% \thanks{Hou Shengren, Peter Palensky and Pedro P. Vergara, are with the Intelligent Electrical Power Grids (IEPG) group, Delft University of Technology, Delft 2628CD, The Netherlands, emails: \{h.shengren, A.fu, P.Palensky, P.P.VergaraBarrios\}@tudeflt.nl.}
% \thanks{T. C. Author is with
% the Electrical Engineering Department, University of Colorado, Boulder, CO
% 80309 USA, on leave from the National Research Institute for Metals,
% Tsukuba, Japan (e-mail: author@nrim.go.jp).}}
\maketitle

\begin{abstract}
The integration of distributed energy resources (DER) has escalated the challenge of voltage magnitude regulation in distribution networks. Traditional model-based approaches, which rely on complex sequential mathematical formulations, struggle to meet real-time operational demands. Deep reinforcement learning (DRL) offers a promising alternative by enabling offline training with distribution network simulators, followed by real-time execution. However, DRL algorithms tend to converge to local optima due to limited exploration efficiency. Additionally, DRL algorithms can not enforce voltage magnitude constraints, leading to potential operational violations when implemented in the distribution network operation. This study addresses these challenges by proposing a novel safe imitation reinforcement learning (IRL) framework that combines IRL and a designed safety layer, aiming to optimize the operation of Energy Storage Systems (ESSs) in active distribution networks. The proposed safe IRL framework comprises two phases: offline training and online execution. During the offline phase, optimal state-action pairs are collected using an NLP solver, guiding the IRL policy iteration. In the online phase, the trained IRL policy's decisions are adjusted by the safety layer to maintain safety and constraint compliance. Simulation results demonstrate the efficacy of Safe IRL in balancing operational efficiency and safety, eliminating voltage violations, and maintaining low operation cost errors across various network sizes, while meeting real-time execution requirements. 
% This research highlights the significance of integrating a safety layer with DRL to enhance system reliability and economic performance in modern power distribution networks.
\end{abstract}


\section{Introduction}

The penetration of renewable energies has pressed emerging challenges to distribution network operators (DSOs) due to the lag in distribution network upgrades, particularly evident in the Netherlands, where the severity of voltage magnitude problems has escalated~\cite{solar2024capacity}. This bottleneck in infrastructure modernization has significantly promoted energy investors to deploy energy storage systems (ESSs) into distribution networks, offering a viable pathway to mitigate voltage magnitude instabilities and enhance the resilience of the distribution network~\cite{van2020low}. In this context, optimizing ESSs dispatch is crucial to ensure voltage regulation while also aiming to minimize operational costs amidst the constraints of an aging network~\cite{aihui2022distributed}.

However, fluctuating prices, varying electricity demands, and uncertainty in renewable generation bring significant challenges in defining the dynamic and sequential optimal operation decisions. Traditional model-based approaches, which rely on predefined forecasts or complex probability functions to manage uncertainties, often struggle with real-time decision-making~\cite{li2023optimal}. As these methods require extensive computational resources, they can be inefficient in adapting to the fast-paced and variable nature of the optimal ESSs dispatch problem.

Deep Reinforcement Learning (DRL) emerges as a promising alternative to traditional model-based approaches, offering a model-free solution that excels in fast-paced, sequential decision-making scenarios~\cite{cao2020drlreview}. DRL has been successfully applied in diverse fields such as game playing, robotics control, and industrial systems, where it transforms operational sequences into Markov Decision Processes (MDPs)~\cite{2022magnetic}. In the context of energy systems tasks, DRL has demonstrated the potential to optimize complex tasks, such as voltage control~\cite{pedro2022_rl_votlage_control} and energy management~\cite{shengren2022performance}, by enabling the DRL algorithms to learn directly from interactions with the built energy system simulator. This capability allows DRL to handle the complexities and uncertainties inherent in distribution networks more effectively~\cite{shengren2022performance}. One of the primary challenges associated with DRL algorithms is low exploration efficiency. The agent requires substantial time to learn due to the need for extensive exploration of the action space~\cite{wang2021multi}. This inefficiency is particularly problematic in scenarios with high-dimensional action spaces, such as controlling multiple ESSs in a distribution network~\cite{hou2023constraint}. This low exploration efficiency consequently leads DRL algorithms to converge prematurely to suboptimal solutions, as fully exploring all possible actions becomes increasingly difficult. For instance, previous research~\cite{hou2023constraint} has shown that DRL algorithms often focus on leveraging only a single ESS that is highly sensitive to voltage magnitude fluctuations, while neglecting the potential flexibility offered by other ESSs. This behavior results in suboptimal performance and prevents the system from fully utilizing the flexibility of multiple ESSs~\cite{mauricio2022eligibility}.

Imitation Learning (IL) offers a complementary approach that can enhance the data efficiency of DRL algorithms~\cite{zheng2022imitation}. IL is a strategy where the learning agent aims to mimic the behavior of an expert by learning from optimal state-action pairs~\cite{goecks2019integrating}. In the context of ESSs dispatch, expert decisions can be derived from solving daily scenarios using commercial solvers, which derive optimal state-action pairs under various scenarios. These pairs provide a high-quality dataset that the RL agent can use to learn desired behaviors without needing to engage in inefficient online exploration~\cite{guo2023sample}. By incorporating IL, the learning process of DRL algorithms is significantly accelerated, as the RL agent starts with a base of optimal actions in different states, thereby reducing the exploration space and focusing on refining strategies that have already proven to be effective. For instance, \cite{cheng2023dispatch} integrated expert demonstrations into the training phase of DRL for real-time dispatch of generation units. Results showed DRL algorithms can achieve faster convergence and improve 2.2\% performance compared to the model-based solution in real-time dispatch tasks. The work in~\cite{dinh2021milpHVAC} applied a Mixed-Integer Linear Programming (MILP)-based IL approach to Heating, Ventilation, and Air Conditioning (HVAC) control. By using IL, a control policy can be trained by imitating the optimal MILP-based decisions, enabling efficient real-time HVAC control without the need for solving complex optimization problems in real-time. In~\cite{xu2022HVAC-BC}, IL is leveraged to accelerate DRL algorithms training for building HVAC control. Results demonstrated DRL algorithms could achieve better control efficiency and effectiveness in managing building HVAC systems. In~\cite{gao2021ILmicrogrid}, an IL-based approach is proposed for online optimal power scheduling of microgrids. The IL-based controller can rapidly adjust power scheduling in real-time, ensuring optimal operation of microgrids under varying conditions by learning from optimal scheduling policies derived from offline optimization models. 

Previous studies have shown that IL or an offline trained IL followed by online DRL fine-tuning can improve the training efficiency and the performance of dispatch policies. However, this combination presents several challenges. First, purely imitation learning-based approaches are highly sensitive to the training dataset, leading to poor generalizability and potentially suboptimal behavior in scenarios that were not part of the training data~\cite{zheng2022imitation}. Second, although online fine-tuning can mitigate this problem, it may also cause a performance collapse due to the state-action distribution shift, where the DRL agent's exploration leads to actions and states that deviate significantly from those seen during the imitation learning phase~\cite{TD3BC}. Third, both of these previous approaches struggle to guarantee the feasibility of the decisions or enforce operational constraints, as they do not explicitly account for feasibility during the imitation learning process~\cite{hou2024MIP-DRL}. In light of these challenges, our contributions are threefold:

\begin{itemize}
    \item We introduce a framework that combines the strengths of DRL algorithms and IL to enhance the training efficiency and dispatch performance of trained algorithms. Moreover, the framework can rigorously enforce operational constraints in distribution networks during the dispatch. This innovative approach addresses the limitations previously identified in these areas.
    \item During the offline training phase, we employ a dual-gradient strategy utilizing both the IL policy and the critic network. This approach stabilizes the training process and expedites learning, effectively overcoming standard DRL algorithms' computational and exploration challenges. 
    \item To guarantee the feasibility of dispatch decisions, the safe layer proposed in our previous paper~\cite{hou2024distflow} is extended to the framework during the online operation. This layer filters out unsafe actions, redirecting them into safer alternatives, thus ensuring the operational feasibility of decisions in scenarios not covered by expert data.
\end{itemize}

% In response to the escalating voltage drop problems exacerbated by the lag in network upgrades, Distribution System Operators (DSO) in the Netherlands (NL) are increasingly turning to the installation of energy storage systems as a strategic intervention. This approach is aimed at directly addressing the challenges posed by inadequate infrastructure modernization, which has left the power distribution network more susceptible to voltage instability. By integrating energy storage solutions, DSOs in NL are not merely reacting to the current dilemmas but are proactively fortifying the grid against voltage fluctuations. This strategy reflects a broader shift towards leveraging advanced technologies to enhance grid reliability and efficiency in the face of growing operational demands and infrastructure constraints. The deployment of energy storage systems thus emerges as a crucial measure for mitigating voltage drop issues, highlighting the role of innovative energy management practices in ensuring the stability and sustainability of the power distribution network in the Netherlands.

\section{Mathematical Formulation of the Optimal ESSs Scheduling Problem}\label{sec_nlp_formulation}
\vspace{-2mm}
The optimal scheduling of ESSs within a distribution network is formulated as a nonlinear programming (NLP) problem, given by \eqref{eq:goal}--\eqref{eq_SOC_cons}. The objective function in \eqref{eq:goal} aims to minimize the total operational cost over the time horizon~${\cal T}$, which includes the costs of importing power from the main grid, dictated by day-ahead market prices $\rho_t$ in EUR/MWh.
\begin{equation}\label{eq:goal}
    \min_{\substack{P^{B}_{m,t}, \forall m \in {\cal B}, \forall t \in {\cal T}}} \left\lbrace  \sum_{t \in {\cal T}} \left[ \rho_{t}\sum_{m \in {\cal N}} \left(P^D_{m,t} + P^{B}_{m,t} - P^{PV}_{m,t}\right)\Delta t \right] \right\rbrace.
\end{equation}
Subject to:
%
% Eq for active power balance
\vspace{-2mm}
\begin{multline} \label{eq:active_power_balance}
 \hspace{-5mm} \sum_{nm \in {\cal L}} P_{nm,t} - \sum_{mn \in {\cal L}} (P_{mn,t} + R_{mn}I_{mn,t}^2) + P_{m,t}^{B} \\ + P_{m,t}^{PV}+ P_{m,t}^{S}= P_{m,t}^{D}  \quad \forall m \in {\cal N}, \forall t \in {\cal T}  
\end{multline}
\vspace{-6mm}
\begin{multline} \label{eq:reactive_power_balance}
 \hspace{-5mm} \sum_{nm \in {\cal L}} Q_{nm,t} - \sum_{mn \in {\cal L}} (Q_{mn,t} + X_{mn}I_{mn,t}^2) + Q_{m,t}^{S} = Q_{m,t}^{D}  \\  \forall m \in {\cal N}, \forall t \in {\cal T}
\end{multline}
\vspace{-6mm}
\begin{multline}
\label{eq_votlage_drop}
 \hspace{-4mm} V_{m,t}^2-V_{n,t}^2=2(R_{mn}P_{mn,t}+X_{mn}Q_{mn,t})+ \\(R_{mn}^2+X_{mn}^2)I_{mn,t}^2 \quad  \forall m,n \in {\cal N}, \forall t \in {\cal T}  
\end{multline}
\vspace{-6mm}
\begin{flalign}
& V_{m,t}^2I_{mn,t}^2=P_{mn,t}^2+Q_{mn,t}^2 & \forall m,n \in {\cal N}, \forall t \in {\cal T} \label{eq_vi=pq} &     
\end{flalign}
\vspace{-6mm}
%
% \begin{multline}
% \hspace{-4mm} SOC_{m,t}^{B}=SOC_{m,t-1}^{B} + \eta^{B}_{m}P_{m,t}^{B}\Delta t/\overline{E}^{B}_{m} \hspace{0.30em} \forall m \in {\cal{B}}, \forall t \in {\cal{T}} \hspace{-0.4em} \label{eq_SOC_cha}
% \end{multline}
%
\begin{multline}
\hspace{-4mm} 
SOC_{m,t}^{B} = SOC_{m,t-1}^{B} + \left\{\begin{array}{ll}
\frac{\eta^{B}_{m,c} P_{m,t}^{B} \Delta t}{\overline{E}^{B}_{m}}, & \text{if } P_{m,t}^{B} > 0 \\
\frac{P_{m,t}^{B} \Delta t}{\eta^{B}_{m,d} \overline{E}^{B}_{m}}, & \text{if } P_{m,t}^{B} < 0
\end{array}\right. \\\quad \forall m \in \cal{B}, \forall t \in \cal{T}
\label{eq_SOC_cha}
\end{multline}

\vspace{-6mm}
\begin{flalign}
& \underline{SOC}_{m}^{B}\leq SOC_{m,t}^{B}\leq\overline{SOC}_{m}^{B} & \forall m \in {\cal{B}}, \forall t \in {\cal{T}} & \label{eq_SOC_cons}\\
& \underline{P}^{B}_{m}\leq P^{B}_{m,t}\leq \overline{P}^{B}_{m} & \forall m \in {\cal B}, \forall t \in {\cal T} & \label{eq_char_disc_cons}\\
& \underline{V}^{2}\leq V_{m,t}^2\leq \overline{V}^{2} & \forall m \in {\cal N}, \forall t \in {\cal T} & \label{eq:voltage_boundary}\\
& 0 \leq I_{mn,t}^2 \leq \overline{I}_{mn}^{2} & \forall mn \in {\cal L}, \forall t \in {\cal T} & \label{eq:limites_corre_5}\\
& P_{m,t}^{S} = Q_{m,t}^{S} = 0 & \forall m \in {\cal N} \backslash \{1\}, \forall t \in {\cal T} & \label{eq:power_not_substation}
\end{flalign}

The distribution network is modeled using the power flow formulation shown in \eqref{eq:active_power_balance}--\eqref{eq_vi=pq} in terms of the active $P_{mn,t}$ power, reactive power $Q_{mn,t}$ and current magnitude $I_{mn,t}$ of lines, and the voltage magnitude $V_{m,t}$ of nodes. Equation in \eqref{eq_SOC_cha} models the dynamics of the ESSs' SOC on the set ${\cal B}$, while \eqref{eq_SOC_cons} enforces the SOC limits. Hereafter, it is assumed that the ESS $m \in {\cal B}$ is connected to node $m$, thus, ${\cal B} \subseteq {\cal N}$. Finally, \eqref{eq_char_disc_cons} enforces the  ESSs discharge/charge operation limits, \eqref{eq:voltage_boundary} and \eqref{eq:limites_corre_5} enforce the voltage magnitude and line current limits, respectively, while \eqref{eq:power_not_substation} enforces that only one node is connected to the substation. Notice that to solve the above-presented NLP formulation, all long-term operational data (e.g., expected PV generation and consumption) must be collected to properly define the ESSs' dispatch decisions, while the power flow formulation must also be considered to enforce the voltage and current magnitude limits. 

\section{MDP Formulation}\label{sec:mdp_formulation}

The above sequential-decision problem can be modeled as a constrained Markov decision process (CMDP), characterized by the tuple $(\mathcal{S},\mathcal{A},\mathcal{P},\mathcal{R},\gamma,\mathcal{C})$. Here, $\mathcal{S}$ denotes the state space which includes observable states of the system,  $\mathcal{A}$ represents the action space of possible control actions, $\mathcal{P}$ is the state transition probability capturing system dynamics, $\mathcal{R}$ is the reward function guiding the policy iteration, $\gamma$ is a discount factor reflecting the importance of future rewards, and $\mathcal{C}$ is a set of constraint functions ensuring operational safety and feasibility. The decision-making follows a policy $\pi(a_t|s_t)$ that selects actions $a_t \in \mathcal{A}$ based on the current state $s_t\in\mathcal{S}$, deriving the system along a trajectory of states, actions, and rewards: $\tau=\left(\boldsymbol{s}_0, \boldsymbol{a}_0, \boldsymbol{s}_1, \boldsymbol{a}_1, \cdots\right)$. The selected actions aimed at maximizing a cumulative reward while adhering to system constraints. 

% This trajectory not only aims to maximize the cumulative reward but also adheres to the system constraints, thereby balancing the objectives of operational efficiency and safety.
%
% \begin{equation}\label{eq:constrained_MDP}
% \begin{aligned}
% &\max _{\pi} J(\pi)=\mathbb{E}_{\tau \sim \pi}\left[\sum_{t=0}^{{\cal T}} \gamma^t r_t\right] \\
% &\text { s.t. } J_{C_i}(\pi) \leq 0, \forall i=1, \ldots, k. \\
% \end{aligned}
% \end{equation}
% %
% \noindent Here, $J_{C_i}(\pi)$ is defined as $J_{C_i}(\pi)=\mathbb{E}_{\tau \sim \pi}\left[\sum_{t=0}^{{\cal T}} \gamma^t C_{i,t}\right] $. A more detailed MDP description of the ESSs optimal scheduling problem is presented below.

The state at time $t$, denoted $s_t$, encapsulates the current operational status of the distribution network and it is defined by the vector: $s_t=[P^{N}_{m,t},V_{m,t}|_{m \in {\cal N}},\rho_t, SOC^{B}_{m,t}|_{m \in {\cal B}},t]$, where $P^{N}_{m,t}=P^{D}_{m,t}-P^{PV}_{m,t}$ represents the net power at node $m$, incoperating both consumption $P^{D}_{m,t}$ and PV generation $P^{PV}_{m,t}$. $V_{m,t}$ is the voltage magnitude at node $m$. $SOC^{B}_{m,t}|_{m \in {\cal B}}$ is the ESS connected to $m_{th}$ node. $t$ is used to indicate which step the agent is in during the whole trajectory. 

% These features can be divided into endogenous and exogenous features. The former includes the PV generation $P^{PV}_{m,t}$ and consumption $P^{D}_{m,t}$, and day-ahead price $\rho_t$, which are independent of the operated actions, while the latter includes the ESSs' SOC $SOC^{B}_{m,t}$, which depends on the agent's action. 

The action $a_t$ at time $t$ involves the dispatch decisions for charging or discharging ESSs, represented by $a_t=[P^{B}_{m,t}|_{m \in {\cal B}}$], where  ${\cal A}$ is a continuous space reflecting the possible charge/discharge power. The transition to the next state $s_{t+1}$ based on the current state $s_{t}$ and action $a_{t}$ is captured by:
\vspace{-2mm}
\begin{multline} \label{eq:transition_function}
  p(S_{t+1},R_{t}|S_t,A_t)= \\ 
    \operatorname{Pr}\left\{S_{t+1}=s_{t+1}, R_{t}=r_{t} \mid S_{t}=s_{t}, A_{t}=a_t\right\}.  
\end{multline}

The transition function $p(\cdot)$ incorporates both the deterministic dynamics of the distribution network and the stochastic nature of demand, PV generation, and market prices.

% \noindent This transition probability function $\cal{P}$ models the endogenous distribution network dynamics, determined by the electrical network itself, and the exogenous uncertainty caused by the PV generation, demand consumption, and the day-ahead price dynamics. In practice, building an accurate mathematical model for such a transition function is not possible. Nevertheless, model-free RL algorithms do not require prior knowledge of function $\cal{P}$ as it can be implicitly learned by interacting with the environment. 

The reward function is designed to reflect the operational cost, defined negatively as: 
\begin{equation}\label{eq_reward}
\hspace{-2mm} {\cal R}_{t}\left(s_{t}, a_{t}\right)= r_t =  - \rho_{t} \left[ \sum_{m \in {\cal N}} \left( P^D_{m,t}+P^{B}_{m,t}-P^{PV}_{m,t}\right)  \right] \Delta t
\end{equation}

To ensure safety and operational feasibility, several constraints are integrated. ESS charging and discharging must not exceed predefined limits ~\eqref{eq_char_disc_cons}. Voltage and current magnitudes must comply with network standards~\eqref{eq:voltage_boundary}, ~\eqref{eq:limites_corre_5}. While constraints on actions and SOC are enforced directly in the policy $\pi$, network constraints are managed indirectly. To handle this, a penalty term is added to the reward function for violations:
\begin{multline}\label{eq:new_penaly_reward}
r_t =  - \rho_{t} \left[ \sum_{m \in {\cal N}} \left( P^D_{m,t}+P^{B}_{m,t}-P^{PV}_{m,t}\right) \right] \Delta t \\ -\sigma \left[  \sum_{m\in\cal{B}} C_{m,t}(V_{m,t})\right] , 
\end{multline}
where $\rho$ is a penalty coefficient, and $C_{m,t}$ is a penalty function for voltage violations, defined to prioritize operational constraints within the learning process. 

\noindent$C_{m,t}$ in \eqref{eq:new_penaly_reward} can be modeled using different functions (e.g., $L_2$ functions). Here, as in~\cite{mauricio2022eligibility}, $C_{m,t}$ is defined as
\begin{equation}
    C_{m,t}=\min \left\{0, \left(\frac{\overline{V}-\underline{V}}{2}-\left|V_0-V_{m, t}\right|\right)\right\}, \forall m \in {\cal B}.
\end{equation}

While the CMDP framework supports the integration of operational constraints, directly applying DRL to optimize ESS scheduling in distribution networks introduces significant challenges. DRL algorithms face high computational demands and often achieve suboptimal policy convergence, struggling to consistently adhere to operational constraints in complex ESS dispatch problems~\cite{shengrenOptimalEnergySystem2023}. Furthermore, after training, DRL agents may fail to enforce these constraints reliably, especially in scenarios that were underrepresented during training. To tackle these issues, we introduce a safe imitation learning framework, which is detailed in the subsequent section.

\section{The Proposed Framework}

\begin{figure}[t!]
   \centering
    \psfrag{R1}[][][0.8]{$[s^{*}, a^{*}]$}
    \psfrag{S1}[][][0.8]{$s_t$}
    \psfrag{A1}[][][0.8]{$a_t$}
    \psfrag{P}[][][0.8]{$\pi$}
    \psfrag{M1}[][][0.8]{$\hat{a}_t$}
\includegraphics[width=1.0\columnwidth]{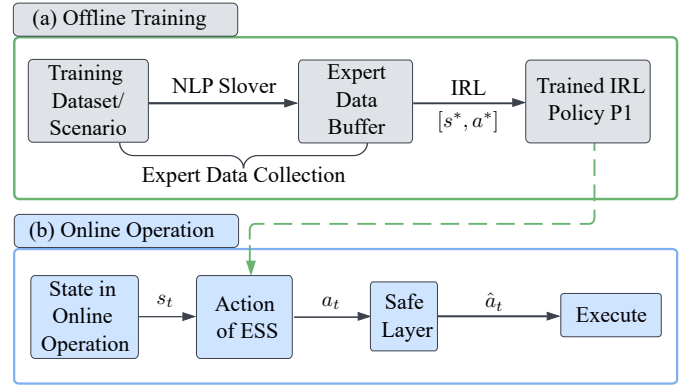}
   \caption{Overall workflow of the proposed framework. The framework is composed of offline and online phases. The offline training is performed once, while the online operation is conducted at each time step $t$.}
   \label{fig_Q_iteration_framework}
   \vspace{-3mm}
\end{figure}

The proposed framework comprises two main phases: offline training and online execution. Initially, during the offline training phase, an expert policy formulated by an NLP solver collects optimal state-action pairs, or \textit{expert data}. This data is used to guide the IRL policy iteration. 

In the online execution phase, the trained IRL policy determines charge/discharge decisions based on the current state. These decisions are then adjusted by the safety layer to ensure strict adherence to operational constraints. This dual-phase approach aims to balance the need for operational efficiency with the essential requirements of safety and constraint compliance.

\subsection{Offline Training Via Imitation Learning}

\subsubsection{Expert Demonstration Data Collection}
The expert demonstration data is crucial for training our IRL framework. Optimal state-action sequences are generated by solving the NLP problem formulated in Section~\ref{sec_nlp_formulation}, capturing a variety of historical scenarios including daily trajectories of renewable generation, load consumption, and price dynamics. This expert policy identifies sequences that minimize operational costs while complying with voltage magnitude constraints, thereby providing a robust dataset for training proposed IRL algorithm.

% Expert demonstration data, embodying optimal state-action sequences across diverse historical scenarios, are gained through the recurrent resolution of a Nonlinear Programming (NLP) problem. Here, the NLP solver acts as the expert policy, delineating optimal strategies while stringently adhering to operational constraints. Specifically, the corresponding scenarios are the daily trajectory for renewable generation, load consumption, and price dynamics, while the actions are the defined sequential charge/discharge decisions of controlled ESSs to minimize the operational cost and enforce the voltage magnitude constraints. Then. the collected expert data are leveraged to train the IRL algorithms, accelerating training and overcoming the performance boundary of the IRL agents.  

\subsubsection{Imitation RL Algorithms}

Reinforcement Learning (RL) emerges as a preeminent strategy for devising policies under uncertainty. Traditional value-based DRL algorithms, such as DQN~\cite{DQN} fail to address the continuous state and action problems. In contrast, Deep Deterministic Policy Gradient (DDPG) algorithm~\cite{DDPG} and it's enhanced counterpart, TD3~\cite{fujimoto_td3_2018}, are capable of handling continuous actions by simultaneously maintaining a policy (actor) $\pi_{\omega}(s_t)$, used to sample actions, and a trained Q-function (critic) $Q_{\theta}(s_t,a_t)$, used to guide the update direction of the policy network. The TD3 algorithm updates the actor-network by
\begin{equation}\label{eq_TD3_policy}
\omega\leftarrow \omega+\nabla_{\omega} \frac{1}{|B|} \sum_{s_t \in B}
\Biggl( \min_{i=1,2} \{ Q_{\theta_i}\left(s_t, \pi_{\omega}(s_t)\right) \} \Biggr),
\end{equation}
while the critic update iteration is defined as
\begin{equation}
    \label{eq_TD3_Q_update}
    \min _{\theta} \sum_{s \in B}\left(r_{t}+\gamma \min_{i=1,2} \{Q_{\theta_{i}^{\text {target}}}\left(s_{t+1}, \pi_{\omega}(s_{t+1})\right) \} - Q_{\theta_{i}}\left(s_{t}, a_{t}\right)\right)^{2}
\end{equation}

% Though it often suffers from low data efficiency and has difficulty in handling continuous states and actions. 
Training the TD3 algorithm to achieve convergence demands extensive interactions between the agents and their environment, a challenge amplified by large state and action spaces. This intensive requirement stems from the necessity for the algorithm to learn from zero. To solve these challenges, the IL approach, specifically behavior cloning (BC), is introduced. BC leverage expert demonstrations to directly map states and actions, thereby significantly enhancing learning efficiency in terms of sample complexity and trading efficiency.
% \textcolor{blue}{introduce behavior cloning}
Given a dataset of state-action pairs $D^{*} = {(s^{*}, a^{*})}$ obtained from expert demonstrations, where $s^{*}$ represents the states observed by the expert and $a^{*}$ represents the corresponding actions taken by the expert policy, the goal of BC is to learn a policy $\pi_{\omega}(s)$ that can generate actions closely approximating the expert's actions for any given state $s$.

The learning process of policy $\pi_{\omega}(s)$ involves adjusting $\omega$ to minimize the difference between the actions predicted by the policy and the expert actions in the dataset. The parameter update for BC is defined as:

\begin{equation}
\label{eq_BC_update}
\omega^* = \arg\min_{\omega} \frac{1}{|B|} \sum_{(s^{*}, a^{*}) \in D^{*}} \left| \pi_{\omega}(s^{*}) - a^{*} \right|^2,
\end{equation}

where $\omega^*$ represents the optimized policy parameters, $B$ is the batch size, $(s^{*}, a^{*})$ are the state-action pairs from the expert demonstrations, and $\pi_{\omega}(s)$ is the policy parameterized by $\omega$.

% This can be formulated as a regression problem with the following loss function:

% \begin{equation}
% \label{eq_BC_loss}
% L(\omega) = \frac{1}{|B|} \sum_{(s_{*}, a_{*}) \in D} | \pi_{\omega}(s_{*}) - a_{*} |^2,
% \end{equation}

% where $L(\omega)$ is the loss function that measures the mean squared error (MSE) between the policy's predicted actions and the expert's actions across all demonstrations in the dataset $D$. The objective is to find the parameters $\omega$ that minimize this loss:

% \begin{equation}
% \omega^* = \arg\min_{\omega} L(\omega).
% \end{equation}

% The parameter update rule, often implemented through gradient descent methods, aims to iteratively adjust $\omega$ in the direction that reduces the loss $L(\omega)$. If using a simple gradient descent approach, the update at each iteration could be expressed as:

% \begin{equation}
% \omega \leftarrow \omega - \alpha \nabla_{\omega} L(\omega),
% \end{equation}

% where $\alpha$ is the learning rate, and $\nabla_{\omega} L(\omega)$ is the gradient of the loss function with respect to the parameters $\omega$.

% By minimizing the loss function $L(\omega)$, 

BC aims to train a policy that can accurately replicate the expert's decision-making process across a wide range of states, thereby leveraging the expert's knowledge to achieve efficient learning especially in environments where exploring through trial and error (as in traditional RL approaches) might be inefficient or infeasible.
However, the main drawback of BC is that if the learner makes a mistake during execution, it may end up in a state completely distinct from the demonstration dataset, which will consequentially lead to error cascading. 

The TD3BC algorithm represents an innovative approach to overcoming the challenges associated with BC, particularly the issue of error cascading when a learner encounters states not covered by the demonstration dataset. TD3BC merges the robustness of DRL with the efficiency of BC for offline training phases. 

The TD3BC algorithm integrates the update mechanisms of both TD3 and BC, formulated as:

\begin{equation}
\label{eq_TD3BC_policy_update}
\omega \leftarrow \omega - \alpha \nabla_{\omega} \left( \lambda_{TD} \frac{1}{|B|} \mathcal{L}_{TD} + \lambda_{BC} \frac{1}{|B|} \mathcal{L}_{BC} \right),
\end{equation}

where: $\mathcal{L}_{TD}$ is the TD loss component, represented by $\left( r_{t} + \gamma \min_{i=1,2} Q_{\theta_{i}^{\text{target}}}(s_{t+1}, \pi_{\omega}(s_{t+1})) - Q_{\theta_{i}}(s_{t}, a_{t}) \right)^2$, $\mathcal{L}_{BC}$ is the BC loss component, represented by $\left| \pi_{\omega}(s^{*}) - a^{*} \right|^2$, $\lambda_{TD}$ and $\lambda_{BC}$ are the weighting coefficients for the TD and BC loss components, respectively, $\alpha$ is the learning rate, $B$ is the batch of transitions sampled from the expert dataset $D^{*}$.

TD3BC innovatively combines the gradients from both the conventional TD loss, used in TD3 for updating the policy and value networks, and an expert loss derived from BC. This dual-gradient approach allows the algorithm to not only learn from the expert demonstrations but also refine its policy via interacting with the environment, as in classical RL, thereby addressing the limitations of each approach when used in isolation.

Despite the TD3BC algorithm's ability to enhance performance and accelerate training, it faces a significant limitation during the online execution phase: it cannot inherently enforce constraints. This limitation stems from the fact that the TD3BC algorithm is trained exclusively on demonstration data, which inherently satisfies operational constraints through the resolution of an NLP problem. Consequently, the algorithm, while effective in replicating demonstrated behaviors, lacks an intrinsic understanding of the safety constraints. This gap in awareness can lead to situations where the actions chosen by the TD3BC-trained agent, when faced with scenarios not covered in the training data, diverge from safe operational bounds, potentially causing serious violations of system constraints.

To address this critical issue and ensure the feasibility and safety of actions during online execution, we propose the integration of a linear safe layer on top of the TD3BC algorithm. This safety layer is designed to function as a regulatory mechanism, adjusting the actions suggested by the TD3BC model to ensure they remain within predefined safety and operational constraints. It acts as a vital check, correcting for the algorithm's lack of direct constraint recognition and ensuring that all actions are compatible with the system's safety requirements. 
% The detail of safe layer formulation and the procedure for safe online execution is explained next. 
% The implementation of this safety layer is crucial for bridging the gap between the theoretical optimality of the TD3BC's decisions and the practical necessity of maintaining system integrity under real-world conditions.

The next section will explain in detail the formulation and operational mechanism of the safe layer, illustrating its role in maintaining both optimized performance and stringent adherence to operational safety constraints.

\subsection{Online Execution with Safe Layer}

The safety layer, introduced in our previous work, leverages a linear approximation of power flow equations to project potentially unsafe actions into a safe operational domain. This projection ensures compliance with system constraints during real-time operation.

\subsubsection{Safe layer formulation}\label{sec:linear-function}

Building on the linear power flow model detailed in [previous paper citation], the safety layer adjusts actions based on a simplified relationship between node voltages and power injections. This linear expression can further be used to derive a direct relationship between the action $\boldsymbol{a}$ vector, corresponding to the dispatch decision of the batteries, i.e. $\boldsymbol{a} =[p^B_{1,t},...p^B_{m,t},...p^B_{|{\cal N}|,t}]$, and $\boldsymbol{v_m^2}$, as next  
\begin{multline}\label{eq:linear_action_pv}
    \boldsymbol{M} \boldsymbol{v^2}=\boldsymbol{M} v_0^2 \mathbf{1}_{|\cal{L}|}+2[\mathrm{D}(\boldsymbol{r_{mn}}) \left(\mathbb{I}-\boldsymbol{T} \boldsymbol{F}^T\right)^{-1} \boldsymbol{T} \boldsymbol{(p^N_{m}-a)}+ \\ 
    \mathrm{D}(\boldsymbol{x_{mn}}) \left(\mathbb{I}-\boldsymbol{T} \boldsymbol{F}^T\right)^{-1} \boldsymbol{T} \boldsymbol{q^N_m}].
\end{multline}

\noindent $\boldsymbol{v_m^2}$ is vector of squared voltage magnitudes at each node, $\boldsymbol{M}$ denotes Matrix relating node voltages to the power injections in the network, $v_0^2$ refers squared voltage magnitude at the source node or substation, $\mathbf{1}_{|\cal{L}|}$ is vector of ones, the size of which matches the number of lines in the network, $\mathrm{D}(\boldsymbol{r_{mn}})$, $\mathrm{D}(\boldsymbol{x_{mn}})$ are diagonal matrices containing line resistances and reactances, respectively, $\boldsymbol{T}$, $\boldsymbol{F}$ are matrices representing the network topology, specifically the connections between nodes. 

The primary focus is on maintaining voltage levels within permissible bounds by modifying the control actions suggested by the RL algorithm. The linear relationship is used to form a mathematical programming problem that finds the closest safe action, minimizing deviations from the initially suggested action while ensuring operational safety:

\begin{equation}
\label{obj_eq_action_projection}
\boldsymbol{\hat{a}}= \underset{\boldsymbol{\hat{a}}}{\arg \min } \frac{1}{2}\left\|\boldsymbol{\hat{a}}-\boldsymbol{a}\right\|^2.
\end{equation}
Subject to:
\vspace{-2mm}
\begin{multline}
\label{eq_voltage_upper_bound}
(\boldsymbol{v_m^2} \mathbf{1}_{|\cal{L}|}+2\boldsymbol{M^{-1}}[\mathrm{D}(\boldsymbol{r_{mn}}) \left(\mathbb{I}-\boldsymbol{T} \boldsymbol{F}^T\right)^{-1} \boldsymbol{T} \boldsymbol{(p^N_{m}-a)}+\\\mathrm{D}(\boldsymbol{x_{mn}}) \left(\mathbb{I}-\boldsymbol{T} \boldsymbol{F}^T\right)^{-1} \boldsymbol{T} \boldsymbol{q^N_{m}}])\leq \overline{v}^2-\epsilon 
\end{multline}
\vspace{-6mm}
\begin{multline}
\label{eq_voltage_lower_bound}
    (\boldsymbol{v_m^2} \mathbf{1}_{|\cal{L}|}+2\boldsymbol{M^{-1}}[\mathrm{D}(\boldsymbol{r_{mn}}) \left(\mathbb{I}-\boldsymbol{T} \boldsymbol{F}^T\right)^{-1} \boldsymbol{T} \boldsymbol{(p^N_{m}-a)}+\\ \mathrm{D}(\boldsymbol{x_{mn}}) \left(\mathbb{I}-\boldsymbol{T} \boldsymbol{F}^T\right)^{-1} \boldsymbol{T} \boldsymbol{q^N_{m}}])\geq \underline{v}^2+\epsilon
\end{multline}

In the above formulation, $\boldsymbol{\hat{a}}$ corresponds to the projected (or safe) action vector. Additionally, due to the error introduced in the linear formulation, a small value $\epsilon$ is added to control the relaxation condition of voltage magnitude limits, following the previous research~\cite{hou2024distflow}.

\subsubsection{Online execution procedure}

The procedure for online execution is illustrated in Algorithm~\ref{Algorithm_online_exeuction}. The trained TD3BC model proposes initial actions based on received states. These actions are then adjusted by the safety layer if they risk violating operational constraints. The algorithm ensures that all actions are safe and reliable before implementation in the distribution network.

\begin{algorithm}[t!]
\label{Algorithm_online_exeuction}
\caption{Online Execution for Safe TD3BC Framework}
Initialize safety layer parameters: $\mathrm{D}(\boldsymbol{r_{mn}}), \mathrm{D}(\boldsymbol{x_{mn}}), \boldsymbol{B}, \boldsymbol{T}, \boldsymbol{M}$.\
Load trained TD3BC model.\
\For{each operational timestep $t$}{
Acquire action $a_t$ from policy $\pi_\omega(s_t)$.\
\If{action $a_t$ risks constraint violation}{
Adjust $a_t$ to $\hat{a}_t$ using safety layer optimization.\
}
Implement action $\hat{a}_t$ in the system.
}
\end{algorithm}

% \begin{figure*}[ht]
%    \centering
%        \psfrag{B1}[][][0.8]{$ \pi_\omega(s)$}
%         \psfrag{B2}[][][0.8]{$Q_\theta(s,\hat{a})$}
%         \psfrag{B3}[][][0.8]{$\nabla_\theta$}
%         \psfrag{B4}[][][0.8]{$\theta\rightarrow\theta_{target}$}
%         \psfrag{B5}[][][0.8]{$\nabla_\omega$}
%         \psfrag{B6}[][][0.8]{$\omega\rightarrow\omega_{target}$}
%         \psfrag{B7}[][][0.8]{$a$}
%         \psfrag{B8}[][][0.8]{$\pi_\omega$}
%         \psfrag{B9}[][][0.8]{$[s,\hat{a}]$}
%         \psfrag{B10}[][][0.8]{$Q_\theta$}
%         \psfrag{C1}[][][0.8]{$\hat{a}$}
%         \psfrag{C2}[][][0.8]{$[s,\hat{a},r,s']$}
%         \psfrag{C3}[][][0.8]{${[s,\hat{a},r,s']}_B$}
%         \psfrag{C4}[][][0.8]{${s}$}
%         \psfrag{A1}[][][0.8]{$\hat{a}=\underset{\hat{a}}{\arg \min } \frac{1}{2}\left\|\hat{a}-a\right\|^2$}
%         \psfrag{A2}[][][0.8]{$s.t.\quad \eqref{eq_voltage_upper_bound},\eqref{eq_voltage_lower_bound}$}
%         \psfrag{A3}[][][0.8]{$ $}
%     \includegraphics[width=1.8\columnwidth]{Q_Iteration.eps}
%    \caption{Architecture of the proposed DF-SRL algorithm displaying the interaction between the actor and critic networks, the safety layer and the interaction process with the environment (distribution network).\textcolor{blue}{Change this figure}}
%    \label{fig_Q_iteration_framework}
% \end{figure*}

\section{Simulation Results}

To evaluate the performance of the proposed Safe TD3BC algorithm, we conduct a comparative analysis with several representative (safe) DRL benchmark algorithms, including TD3 algorithm, Safe TD3 algorithm and TD3BC algorithm. In addition, a centralized model-based approach, an NLP formulation~\cite{aihui2022distributed} with perfect forecast information is counted as the global optimality. We first evaluate the performance of the proposed Safe TD3BC algorithm in a 34-node distribution network and then scalability analysis is conducted in diverse sizes of distribution network cases (18-node, 69-node, and 124-node). All these distribution network environments are provided in the open-sourced package~\cite{RL-ADN}.
The parameters for different DRL algorithms and cases are summarized in Table~\ref{table:hyperparamters}. TD3, Safe TD3  algorithms are trained with the same hyperparameters as safe TD3BC algorithms. The parameters of the implemented safe layer follow our previous research~\cite{hou2024distflow}.
Note that while all the DRL benchmark algorithms can make decisions only using current information and achieve online operation, the solution obtained by the NLP formulation requires complete information of the foreseen control period. To train and assess the performance of the DRL benchmark algorithms, we employ validation metrics based on the negative value of total used active power, as denoted in~\eqref{eq_reward}, and the voltage magnitude violation penalty as specified in~\eqref{eq:new_penaly_reward}, counted as the cost of the voltage magnitude violation. These metrics effectively gauge the operational efficiency and constraint adherence of each algorithm.
% All implemented algorithms and their hyperparameters are available in~\cite{Shengrencode}.

\begin{table}[]
\caption{Summary - Parameters for DRL algorithms and the environment}\label{table:hyperparamters}
\centering
\scalebox{0.90}{
\begin{tabular}{cc}
\toprule
\multirow{4}{*}{TD3, Safe TD3} & $\gamma=0.995$\\
                           & $\text{Optimizer adopts Adam}$\\
                           & \text{Learning rate is} $6e-4$\\
                           & \text{Batch size is} $512$ \\
                           &
                           \text{Replay buffer size is} $4e5$ \\
                           \hline
\multirow{4}{*}{TD3BC} & $\gamma=0.995$\\
                           & $\text{Optimizer adopts Adam}$\\
                           & \text{Learning rate is} $6e-4$\\
                           & \text{Batch size is} $512$, \text{Replay Buffer is} $4e5$\\
                           & $\lambda_{BC}=0.5$, $\lambda_{TD}=0.5$\\
                           \hline
\multirow{4}{*}{Safe TD3BC} & $\gamma=0.995$\\
                           & $\text{Optimizer adopts Adam}$\\
                           & \text{Learning rate is} $6e-4$\\
                           & \text{Batch size is} $512$, \text{Replay Buffer is} $4e5$ \\
                           &
                           $\lambda_{BC}=0.5$, $\lambda_{TD}=0.5$ \\
                           \hline
Reward                     & $\sigma=400$ \\ \hline

\multirow{2}{*}{ESSs}& $\overline{P}^B=150kW,\underline{P}^B=-150kW$, \\
&$\overline{SOC}^B=0.8,\underline{SOC}^B=0.2,\eta^B_{c}/\eta^B_{d}=0.98$ \\ \hline
\bottomrule
\vspace{-6mm}
\end{tabular}\label{tab:table_summ}}
\end{table}

\subsection{Performance on Training Set}

\begin{table}[t]
\centering
\caption{Performance and training time of algorithms on simulated 34-node distribution network.}
\label{tab_compare_train_performance}
\scalebox{0.9}{
\begin{tabular}{cccc}
\hline
\multicolumn{1}{l}{Algorithms} & \begin{tabular}[c]{@{}c@{}}Training Time $[h]$\end{tabular} & \begin{tabular}[c]{@{}c@{}}Converged \\ Reward [-]\end{tabular} & \begin{tabular}[c]{@{}c@{}}Violations [-]\end{tabular} \\ \hline
TD3        & 4.3           & 1.7$\mypm$0.1\              & -0.9$\mypm$0.1\       \\
Safe TD3   & 17.1          & 2.4$\mypm$0.3\              & -1.6$\mypm$0.8\      \\
TD3BC      & 0.9           & 4.9$\mypm$0.5\              & -7.7$\mypm$1.5\       \\
Safe TD3BC & 0.9           & 4.5$\mypm$0.1\              & 0          \\ \hline
\end{tabular}}
\vspace{-2mm}
\end{table}

Table~\ref{tab_compare_train_performance} presents the performance and training time of Safe TD3BC and benchmark algorithms applied to a simulated 34-node distribution network. The key metrics evaluated are training time (in hours), converged reward, and violations. A higher converged reward indicates better algorithmic performance, while negative values in the violations column signify undesirable constraint breaches.

The TD3 algorithm achieves a moderate converged reward of 1.7 ± 0.1, with some violations recorded at -0.9 ± 0.1. In contrast, the Safe TD3 algorithm improves the performance with a higher converged reward of 2.4 ± 0.3. However, this improvement comes at the cost of increased violations (-1.6 ± 0.8), suggesting that while the Safe TD3 algorithm optimizes the reward better than the TD3 algorithm, it struggles to adhere to constraints effectively. The TD3BC algorithm emerges as the top performer in terms of the converged reward, achieving the highest value of 4.9 ± 0.5. TD3BC effectively harnesses the potential of offline data, enabling the algorithm to quickly adapt to profitable strategies collected in past operations. Consequently, the TD3BC algorithm shows a marked improvement in performance metrics, capitalizing on the accumulated knowledge embedded in the dataset to optimize actions more efficiently than TD3 and Safe TD3 algorithms. Nevertheless, the TD3BC algorithm caused the most severe violations recorded at -7.7 ± 1.5, indicating that the algorithm does not sufficiently enforce safety constraints. This high performance coupled with significant violations highlights a critical failure of the TD3BC algorithm to maintain safe operations. In contrast, the Safe TD3BC algorithm presents a more balanced approach, combining high performance with strict constraint enforcement. The Safe TD3BC algorithm achieves a converged reward of 4.5 ± 0.1, slightly lower than TD3BC without any voltage magnitude violations. These results suggest that the Safe TD3BC algorithm effectively optimizes performance while strictly enforcing safety constraints, making it a robust choice for applications requiring high reliability and safety.

% \textcolor{red}{We trained a lot more algorithms, in here I only put these 4, the reason is that in the next part comparing the detail result, we justed showed these 4 algorithms because of the space. Do you think this is suitable or not}

\subsection{Dispatch Decision Comparision on Testing Dataset}
\begin{figure*}[!htbp]
    \centering
    \includegraphics[width=1.6\columnwidth]{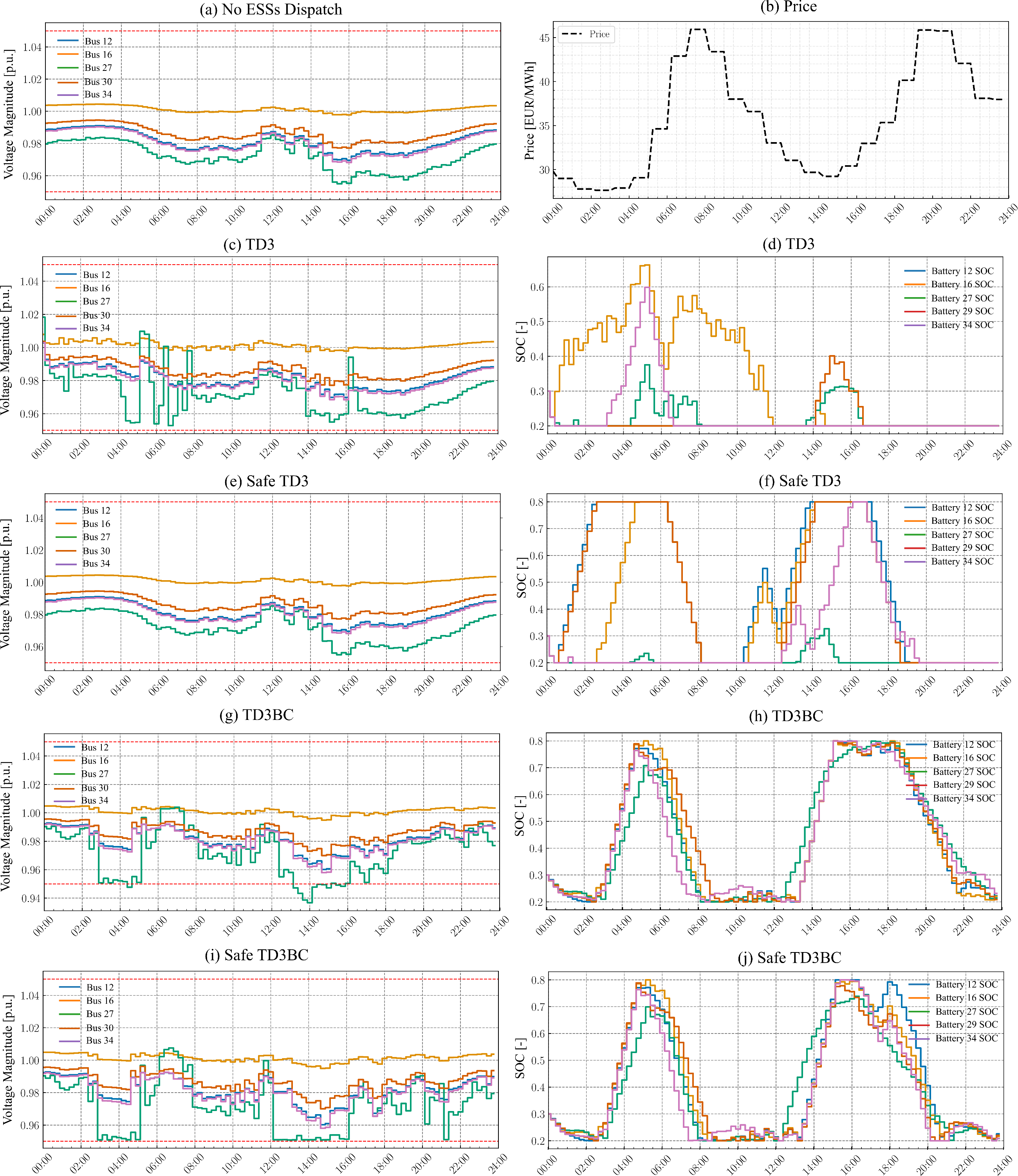}
        \caption{(a): Voltage magnitude for nodes in which the ESSs are connected, disregarding their operation. (b): Price in \texteuro/MWh. Voltage magnitude ((c), (e) (g)) in which the ESSs are connected and SOC of ESSs ((d), (f), (h)), after executing the dispatch decisions.}
    \label{fig_soc_decisions}
\end{figure*}

Fig.~\ref{fig_soc_decisions} displays the voltage magnitude of the nodes in which the ESSs are connected and the SOC of each ESS during a typical day in the test dataset. Results shown in Fig.~\ref{fig_soc_decisions} are obtained after using the dispatch decisions provided by the TD3, Safe TD3, TD3BC, and Safe TD3BC algorithms. Fig.~\ref{fig_soc_decisions}$(a)$ shows the voltage magnitude of the nodes in which the ESSs are connected, but in this case, disregarding their operation (i.e., ESSs are neither charging nor discharging), while Fig.~\ref{fig_soc_decisions}$(b)$ shows the day-ahead electricity price of that test day.

The TD3 algorithm optimizes ESSs operations by responding to price signals, as shown in Fig.~\ref{fig_soc_decisions}$(d)$. This enables it to define charging and discharging decisions of ESSs to maximize profit margins. However, the TD3 algorithm does not fully leverage the potential flexibility of all ESSs. For instance, the TD3 algorithm primarily dispatches the ESS connected to Bus 16, largely ignoring the flexibility offered by ESSs connected with other nodes. Additionally, the TD3 algorithm fails to leverage the evening price peaks, indicating convergence to a local optimum. In contrast, the TD3BC algorithm demonstrates a more aggressive strategy, as shown in Fig.~\ref{fig_soc_decisions}$(h)$. It exploits ESSs flexibility to a greater extent by scheduling ESSs operations aggressively to capitalize on favorable price periods. The TD3BC algorithm maximizes economic gains but at the cost of frequent voltage violations, especially notable during low price periods such as between 02:00 and 04:00, 12:00 and 14:00, causing serious voltage magnitude drops for node 27. 

The Safe TD3BC algorithm eliminates the risk of voltage magnitude violations while fully leveraging the flexibility provided by ESSs connected to all nodes. The safety layer actively adjusts the decisions of the TD3BC algorithm, projecting potentially unsafe actions into safe domains. These modifications, which follow the principle of minimizing the Euclidean distance to the original actions, are designed to prevent safety breaches while maintaining the integrity of operational goals. Fig.~\ref{fig_soc_decisions}$(j)$ shows how the Safe TD3BC algorithm manages the SOCs effectively without causing voltage magnitude violations, as evident from the stable voltage magnitude in Fig.~\ref{fig_soc_decisions}$(i)$. While the safety layer introduces some trade-offs in terms of reduced economic performance due to necessary adjustments to ensure safety, the overall impact is profoundly positive. Safe TD3BC substantially enhances system reliability, effectively eliminating voltage magnitude violations without significantly compromising on economic benefits.

\begin{figure}[!htbp]
    \centering
\includegraphics[width=1.0\columnwidth]{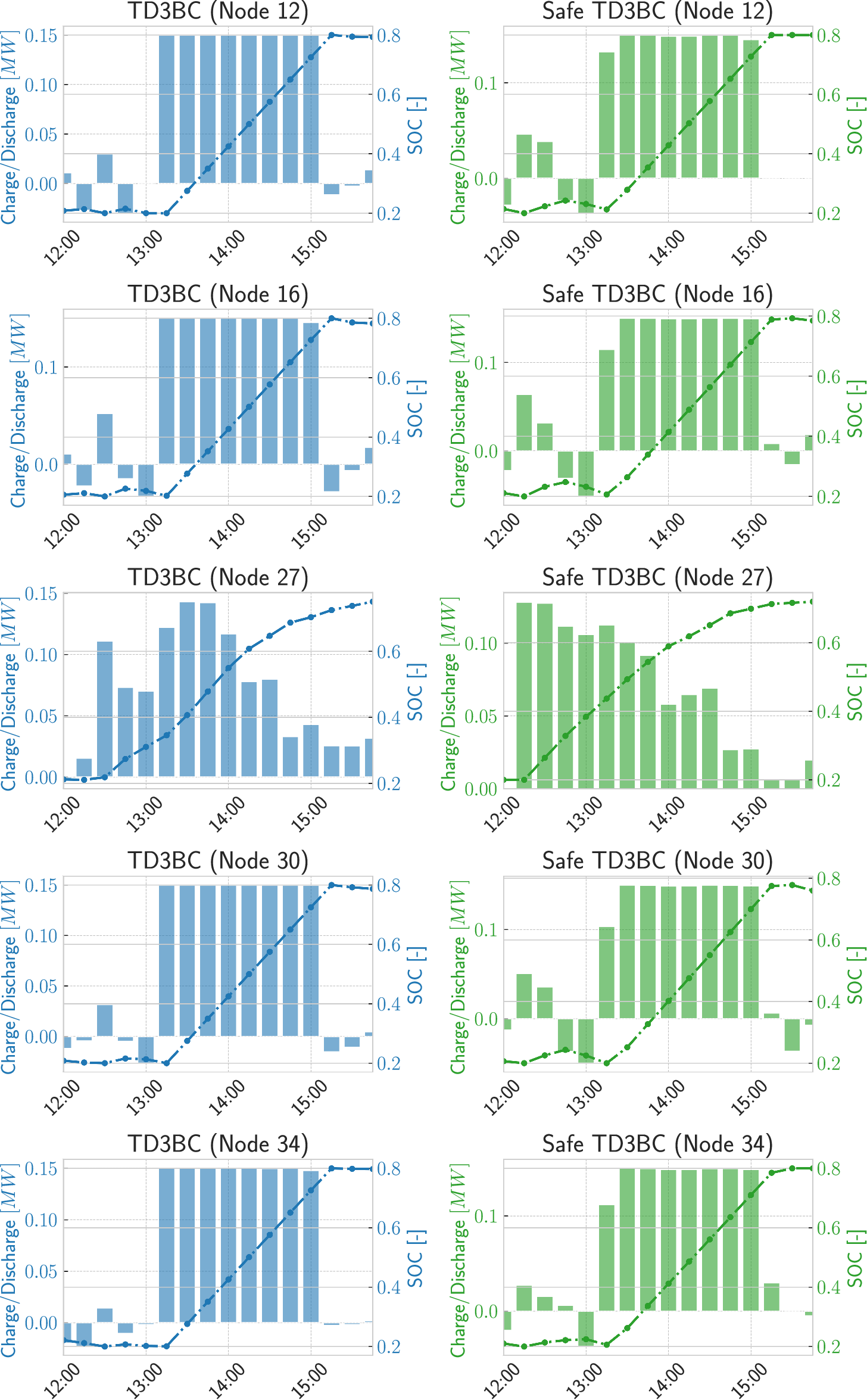}
    \caption{ESSs dispatch patterns between 12:00-16:00, conducted by the TD3BC and SafeTD3BC algorithms.}
     \label{fig_detail_charge}
\end{figure}

Fig.~\ref{fig_detail_charge} displays the detailed charge and discharge patterns for the TD3BC and Safe TD3BC algorithms across nodes 12, 16, 27, 30, and 34 between 12:00 and 16:00. The TD3BC algorithm, depicted in the left column of Fig.~\ref{fig_detail_charge}, demonstrates a clear strategy of aggressive charging and discharging. For instance, at node 12, The TD3BC algorithm charges the ESSs up to 0.15 MW at 13:15, significantly increasing the SOC. Similar patterns are observed at nodes 16, 27, 30, and 34, where the TD3BC algorithm aims to capitalize on this price period. However, this aggressive strategy leads to serious voltage violations. The aggressive charging caused a significant voltage drop in node 27, leading to voltage violations. In contrast, the Safe TD3BC algorithm, shown in the right column of Fig.~\ref{fig_detail_charge}, incorporates a safety layer that modifies the charge and discharge decisions to avoid voltage magnitude violations. The Safe TD3BC algorithm still engages in charging and discharging to maximize operational benefits but also guarantee the feasibility of voltage magnitude constraints. Instead of fully charging at 13:15, the Safe TD3BC algorithm maintains a more moderated charging pattern, ensuring the SOC gradually increases without causing voltage magnitude violations. This pattern also modified the decision at 13:00, where the Safe TD3BC algorithm adjusts the charging strategy to prevent the issues observed with the TD3BC algorithm. Across all nodes, the Safe TD3BC algorithm consistently ensures that the SOC increases in a controlled manner. This careful adjustment of charging and discharging schedules highlights the capability of the Safe TD3BC algorithm to balance economic benefits with strict adherence to safety constraints.

\subsection{Scalability Analysis}

\begin{table*}[h!]
\centering
\caption{Scalability Analysis of Algorithms on Different Network Sizes}
\label{tab_scala}
\scalebox{0.9}{
\begin{tabular}{ccccccc}
\toprule
\multirow{2}{*}{Nodes} & \multirow{2}{*}{Algorithm} & \multirow{2}{*}{\begin{tabular}[c]{@{}c@{}}Exper Data \\ Collection Time [h]\end{tabular}} & \multirow{2}{*}{\begin{tabular}[c]{@{}c@{}}Training \\ Time [h]\end{tabular}} & \multirow{2}{*}{\begin{tabular}[c]{@{}c@{}}Exec. \\ Time [s]\end{tabular}} & \multirow{2}{*}{\begin{tabular}[c]{@{}c@{}}Operation Cost \\ Error (\%)\end{tabular}} & \multirow{2}{*}{\begin{tabular}[c]{@{}c@{}}Voltage Magn. \\ Violations counts [-]\end{tabular}} \\
 & & & & & & \\ \hline
\multirow{4}{*}{18}  & TD3        & - & 4 & 15$\mypm$0.1 & 33.2$\mypm$1.1 & 14$\mypm$2 \\
    & Safe TD3   & - & 12 & 20$\mypm$1 & 15.7$\mypm$0.8 & 6$\mypm$1 \\
    & TD3BC      & 0.5 & 0.7 & 15$\mypm$0.1 & 3$\mypm$0.5 & 45$\mypm$11 \\
    & Safe TD3BC & 0.5 & 0.7 & 20$\mypm$0.7 & 7$\mypm$0.4 & 0 \\
    \hline
\multirow{4}{*}{34}  & TD3        & - & 4 & 15$\mypm$0.1 & 35.9$\mypm$0.9 & 19$\mypm$4 \\
    & Safe TD3   & - & 17 & 25$\mypm$1 & 19.8$\mypm$1.4 &25$\mypm$7 \\
    & TD3BC      & 1.7 & 0.9 & 15$\mypm$0.1 & 6$\mypm$2.5 & 98$\mypm$25 \\
    & Safe TD3BC & 1.7 & 0.9 & 22$\mypm$0.5 & 10$\mypm$0.1 & 0 \\
    \hline
\multirow{4}{*}{69}  & TD3        & - & 4.9 & 15$\mypm$0.1 & 28.5$\mypm$0.4 & 35.1$\mypm$2 \\
    & Safe TD3   & - & 25 & 37$\mypm$2 & 39.9$\mypm$5.6 &277$\mypm$87 \\
    & TD3BC      & 2.5 & 1.5 & 15$\mypm$0.1 & 6.9$\mypm$0.3 & 286$\mypm$35 \\
    & Safe TD3BC & 2.5 & 1.5 & 28$\mypm$0.5 & 9.5$\mypm$0.5 & 0 \\
    \hline
\multirow{4}{*}{124} & TD3        & - & 9 & 15 $\mypm$0.1 & 49.8$\mypm$0.4 & 33$\mypm$2 \\
    & Safe TD3   & - & 43 & 75$\mypm$13 & 105$\mypm$10.1 & 958$\mypm$109 \\
    & TD3BC      & 9 & 2.9 & 15$\mypm$0.1 & 11.5$\mypm$0.7 & 705$\mypm$15 \\
    & Safe TD3BC & 9 & 2.9 & 36$\mypm$1 & 15.9$\mypm$2.2 & 0 \\
    \hline
\bottomrule
\end{tabular}}
\end{table*}

Table~\ref{tab_scala} provides a comprehensive overview of the scalability and performance of four different algorithms (TD3, Safe TD3, TD3BC, and Safe TD3BC) across various network sizes (18, 34, 69, 124 nodes). TD3BC and Safe TD3BC algorithms show better performance compared to TD3 and Safe TD3 algorithms, primarily due to their use of expert data. For smaller networks (18 nodes), the TD3BC algorithm achieves an operation cost error of 3 ± 0.5\%, significantly lower than the performance of the TD3 algorithm, 33.2 ± 1.1\%. However, the TD3BC algorithm fails to enforce safety constraints, resulting in 45 ± 11 violations for the 18-node network. As network size increases, operation cost error of the TD3BC algorithm rises to 11.5 ± 0.7\% for the 124-node network, along with a substantial increase in violations (705 ± 15).

In contrast, the Safe TD3BC algorithm consistently maintains low operation cost errors and zero violations across all network sizes. For instance, the Safe TD3BC algorithm has an operation cost error of 7 ± 0.4\% for 18 nodes network and 15.9 ± 2.2\% for 124 nodes network, without any voltage magnitude violations. This demonstrates the ability of the Safe TD3BC algorithm to balance performance and safety effectively.

TD3 and Safe TD3 algorithms, although not requiring expert data, struggle with constraint enforcement. The TD3 algorithm shows a high number of violations across all network sizes, with 14 ± 2 violations for 18 nodes network and 33 ± 2 for 124 nodes network. The Safe TD3 algorithm performs worse than the TD3 algorithm in larger networks, showing 277 ± 87 violations for 69 nodes and 958 ± 109 for 124 nodes. This indicates that Safe TD3 is not effective in enforcing safety constraints of larger networks.

All algorithms meet real-time requirements, with execution times remaining relatively stable across different network sizes. TD3 and TD3BC algorithms maintain execution times of approximately 15 seconds, while Safe TD3 and Safe TD3BC algorithms have slightly higher execution times due to the additional computations required for enforcing safety constraints. For example, the Safe TD3 algorithm requires 75 ± 13 seconds for a 124-node network, while the Safe TD3BC algorithm requires 36 ± 1 seconds. The lower execution time of the Safe TD3BC algorithm compared to the Safe TD3 algorithm can be attributed to two factors: fewer activation of the safe layer in Safe TD3BC and easier projection of actions, as most actions in Safe TD3BC lie within the boundary.

The preparation of expert data can be time-consuming, as it involves repeatedly solving large-scale optimization problems. This is an offline process and does not impact real-time performance. TD3BC and Safe TD3BC algorithms require less than 3 hours to collect expert data for smaller networks, but this increases to 9 hours for the 124-node network. Training times vary significantly across algorithms and network sizes. Safe TD3 consistently requires more training time than TD3 due to the computational effort involved in ensuring safety constraints. For instance, the Safe TD3 algorithm requires 43 hours to train on a 124-node network, compared to 9 hours for TD3 algorithm. TD3BC and Safe TD3BC algorithms have shorter training times, with the Safe TD3BC algorithm maintaining a training time of 2.9 hours even for the largest network. The extended training time for the Safe TD3 algorithm is primarily due to the frequent activation of the safe layer during environment interactions, which consumes substantial computational resources.

% \textcolor{red}{scalability analysis done}

\section{Discussion}
% \textcolor{blue}{}

DRL algorithms are designed to optimize decision-making based on the rewards obtained through interactions with the environment. A critical component of their learning process is the exploration of the action space to discover strategies that maximize long-term rewards. However, our findings suggest that standard DRL algorithms often struggle with efficient exploration, particularly in complex operational contexts such as the dispatch of ESSs in distribution networks. One of the key issues observed is that the TD3 algorithm tends to fully dispatch the ESS connected to the node experiencing voltage magnitude issues, while neglecting the dispatch of other ESSs. This behavior leads the algorithm to converge to local optima, rather than exploring more globally optimal strategies. The underlying reason is that DRL algorithms inherently lack mechanisms to sufficiently diversify their exploration, especially in environments characterized by high-dimensional continuous action spaces or intricate reward structures. As a result, once the algorithm identifies a reasonably effective solution, it tends to exploit this solution excessively, foregoing further exploration of potentially superior alternatives~\cite{hou2024MIP-DRL}.

Moreover, while introducing a soft penalty component into the reward structure to enforce operational constraints can help mitigate unsafe actions, it often comes at the cost of overall performance. DRL algorithms must be sensitive to these penalties to avoid violating constraints, which inadvertently shifts the learning focus toward avoiding dangerous actions rather than improving overall performance. This heightened sensitivity to penalties amplifies the significance of actions that frequently lead to violations, causing the algorithm to overemphasize the avoidance of those specific actions. Consequently, the DRL agent may ignore other aspects of the action space that could contribute to better performance, ultimately leading to premature convergence to a local optimum. While the soft penalty approach increases safety, it does so by reducing the algorithm's ability to explore and optimize across other dimensions of the action space, thereby limiting the potential for achieving higher performance.

The TD3BC algorithm integrates BC to accelerate the learning process and improve the action quality by guiding the policy towards historically expert actions. This method effectively harnesses the potential of offline data, enabling the algorithm to quickly adapt to profitable strategies collected in past operations. Consequently, the TD3BC algorithm shows a marked improvement in performance metrics, capitalizing on the accumulated knowledge embedded in the dataset to optimize actions more efficiently than its standard counterpart. Nevertheless, the TD3BC algorithm introduces significant risks related to safety compliance, primarily due to its unawareness of the safety constraints. The expert training dataset cannot encompass all possible real-world scenarios, and when the real-world conditions deviate from the training scenarios, the model may fail to recognize or avoid actions that could lead to operational hazards, such as voltage magnitude violations. This limitation highlights a critical weakness in the TD3BC algorithm: while it can improve performance, it cannot ensure safety under unforeseen conditions.

To address this shortcoming, we propose the Safe TD3BC algorithm, which builds upon the strengths of imitation learning while incorporating mechanisms to guarantee safety. The Safe TD3BC framework not only retains the performance improvements of TD3BC by efficiently dispatching all ESSs, but also introduces a layer of safety that ensures compliance with operational constraints, even in scenarios not covered by the training data. By filtering unsafe actions and providing safer alternatives, the Safe TD3BC algorithm significantly enhances both the performance and safety of ESS dispatch, thus overcoming the limitations of the original TD3BC approach.

\section{Conclusion}

The comprehensive analysis reveals that the Safe TD3BC algorithm excels in balancing operational efficiency and safety in active distribution networks. Unlike the TD3 and Safe TD3 algorithms, which struggle to exploit the full flexibility of energy storage systems (ESS) and often converge to local optima, Safe TD3BC leverages expert data to enhance performance while maintaining strict safety constraints. The TD3BC algorithm, although effective in maximizing economic gains, frequently leads to significant voltage violations due to its aggressive dispatch strategy, especially during low-price periods. In contrast, Safe TD3BC incorporates a safety layer that modifies potentially unsafe actions into safe ones, thereby preventing voltage violations without considerably sacrificing economic benefits. In summary, Safe TD3BC is a robust solution, combining high operational efficiency with strict safety enforcement, making it superior to other DRL algorithms like TD3, Safe TD3, and TD3BC, which either fail to maximize performance or enforce safety constraints effectively.
\small 
\bibliographystyle{IEEEtran}
\bibliography{citation} 
\vspace{-2mm}

% \section*{Acknowledgment}

% The preferred spelling of the word ``acknowledgment'' in American English is
% without an ``e'' after the ``g.'' Use the singular heading even if you have
% many acknowledgments. Avoid expressions such as ``One of us (S.B.A.) would
% like to thank $\ldots$ .'' Instead, write ``F. A. Author thanks $\ldots$ .'' In most
% cases, sponsor and financial support acknowledgments are placed in the
% unnumbered footnote on the first page, not here.

\end{document}